\documentclass[aps,prb,floatfix,preprint,showpacs]{revtex4}
\usepackage{graphicx}

\begin{document}

\title{Tunneling Anisotropic Spin Polarization in lateral (Ga,Mn)As/GaAs spin Esaki diode devices}

\author{A. Einwanger, M. Ciorga, U. Wurstbauer, D. Schuh, W. Wegscheider and D. Weiss}
\affiliation{Experimentelle und Angewandte
Physik, University of Regensburg, D-93040 Regensburg, Germany.}

\begin{abstract}
We report here on anisotropy of spin polarization obtained in lateral all-semiconductor all-electrical spin injection devices, employing $p^{+}-$(Ga,Mn)As/$n^{{+}}-$GaAs Esaki diode structures as spin aligning contacts, resulting from the dependence of the efficiency of spin tunneling on the orientation of spins with respect to different crystallographic directions. We observed an in-plane anisotropy of $~8\%$ in case of spins oriented either along $[1\bar{1}0]$ or $[110]$  directions and $~25\%$ anisotropy between in-plane and perpendicular-to-plane orientation of spins. 
\end{abstract}


\maketitle

Realizing novel functional devices based on spins in semiconductors requires efficient spin injection of spin-polarized electrons into semiconducting material and subsequent detection of the resulting spin accumulation.\cite{zutic} One of the very promising materials to utilize  as a source of spin-polarized carriers in such devices is the III-V ferromagnetic semiconductor (Ga,Mn)As that can be grown epitaxially on GaAs-based heterostructures.\cite{ohno} Employing an Esaki-Zener  diode structure (Ga,Mn)As/$n^{+}$-GaAs allows using that $p$-type material as an efficient source of spin-polarized electrons.\cite{kohda2001,johnston,vandorpe2004} Under small negative bias electrons from the valence band of (Ga,Mn)As can tunnel to the conduction band of GaAs and spin injection takes place. Under positive bias tunneling in reverse direction occurs and spin polarization in GaAs is established due to extraction of majority spins from GaAs. As was shown experimentally and supported theoretically the spin polarization can be relatively high at low biases and then decreases with increasing bias.\cite{vandorpe2005,kohda2006} 

In a recent report\cite{ciorga2009} we have presented a lateral device, based on (Ga,Mn)As/GaAs structure, with a successful implementation of an efficient all-electrical spin injection and detection scheme. We have measured a relatively high spin injection efficiency $P$ of $\approx 50\%$ for low bias currents $|I|\leq 10\mu A$. In this study we test such a device for the dependence of the obtained spin injection efficiency on the orientation of injected spins. As recently discovered, the tunnel resistance in structures with a single (Ga,Mn)As ferromagnetic layer depends on the relative orientation of magnetization in that layer with respect to the direction of crystalographic axes leading to an effect called tunneling anisotropic magnetoresistance (TAMR).\cite{gould} By analogy, the anisotropy in related polarization of tunneling current could be described as tunneling anisotropic spin polarisation (TASP).\cite{fabian}    

The experiments were performed on devices of a similar type as the one used in Ref.~\onlinecite{ciorga2009}. The schematic of the device is shown in Fig. 1(a). The sample features four magnetic Esaki diode contacts (2--5) and two non-magnetic contacts (1, 6) to the transport channel. The size of the magnetic contacts is $1\times50 \mu m$ and the spacing between their centers is 6, 7, and 6 $\mu$m between pairs 3--2, 4--3, and 5--4, respectively. Esaki diodes consist of 20 nm of Ga$_{0.95}$Mn$_{0.05}$ and 8 nm of $n^{+}$-GaAs, with $n^{+}=6\times10^{18} $cm$^{-3}$. The transport channel is a 250 nm thick $n-$GaAs layer with $n=6\times10^{16} $cm$^{-3}$. Between the diode and the channel a 15 nm thick GaAs transition $n^{+}\rightarrow n$ layer is also used. More details on the used wafer and fabrication process can be found in Ref.~\onlinecite{ciorga2009}. The measurements were done in a non-local geometry\cite{johnson}, i.e., with no current flowing in a detector circuit. The resistance of the injecting contacts was also monitored by measuring the voltage $V_{inj}$. As there is no current flowing between contacts 2--6, $V_{inj}$ reflects only voltage drop across the interface of the injecting contact 2, without a contribution from the resistance of the lateral transport channel. A non-equilibrium spin accumulation induced in $n-$GaAs underneath the injector and diffusing in either direction of this contact results in the non-local voltage $V^{nl}$, measured by the detector at a distance $L$ from the injector, and is given by\cite{fabian}  
\begin{equation}
	V^{nl}=\pm(P_{inj}P_{det}I\lambda_{sf}\rho_{N}/2S)\exp(-L/\lambda_{sf})+V_{offset}
\end{equation}
where $I$ is a bias current, $\rho_{N}$, $\lambda_{sf}$ , $S$ are, respectively, resistivity, spin diffusion length and the cross-section area of the non-magnetic channel. $P_{inj(det)}$ is the spin injection efficiency of the injector (detector) contact and expresses the polarization of the current injected at the respective contact. The $\pm$ sign corresponds to a parallel (anti-parallel) configuration of magnetizations in ferromagnetic electrodes and $V_{offset}$ is the background offset value observed in many non-local spin-valve (SV) measurements\cite{lou,jedema}, the origin of which is still not fully understood. To make the analysis of the experimental data simpler we limit here our discussion to the case of low bias values, for which $P_{inj}=P_{det}=P$ holds\cite{ciorga2009}. We present then mostly the results of ac lock-in measurements with low excitation current, typically $I=5\mu A$. The use of ac technique allowed to increase the signal-to-noise ratio for low-level excitation and to minimize the background in the signal, as the latter appears to be roughly symmetric with the bias current. The downside of ac measurement is however the loss of information about the difference between injection and extraction of spins.

\begin{figure}
\label{f1}
\begin{center}
\includegraphics[width=0.7\columnwidth,clip]{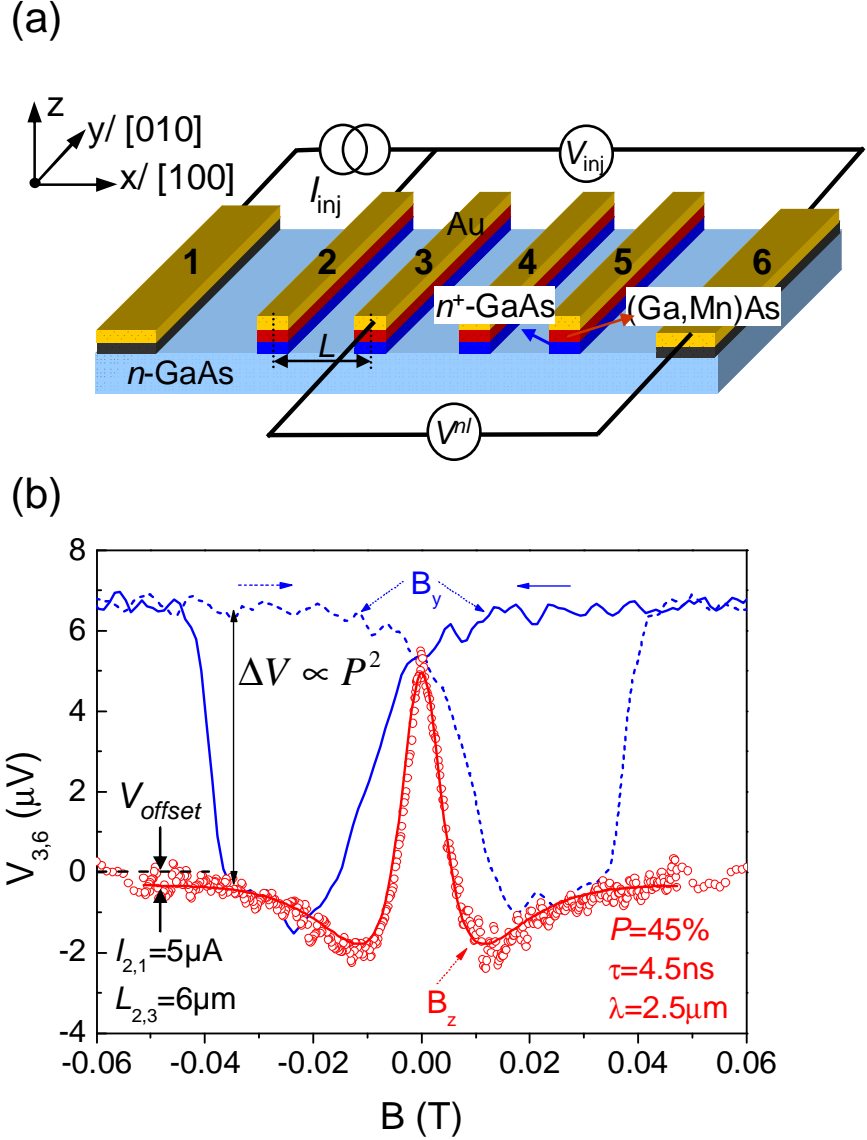}
\caption{(color online) (a) schematics of the experimental device; (b) typical dependence of the non-local signal on in-plane (blue curves) and out-of-plane (red curves) magnetic field. Symbols in the latter case indicate experiment and solid line is a theoretical fit with fitting parameters also shown}
\end{center}
\end{figure}
In Fig. 1(b) we show typical behavior of the detector signal in both out-of-plane magnetic field $B_{z}$ along [001] crystallographic direction and in-plane field $B_{y}$ along [010] direction. In the former case one observes well-known oscillations of the signal and its suppression as a result of spin precession and dephasing in the magnetic field transversal to the initial orientation of spins due to Hanle effect\cite{fabian}. As at sufficiently high field $B_{z}\approx 0.05T$ the spin signal is equal to zero we can extract the offset background signal as $V_{offset}\approx-0.4\mu V$. By modeling the data with theoretical curves\cite{ciorga2009,fabian} we obtain spin injection efficiency of $P\approx 45\%$, spin relaxation time of 4.5 ns and spin diffusion length of 2.5 $\mu$m, i.e., values consistent with the earlier report\cite{ciorga2009}. In the in-plane-field we observe switching in the SV-like fashion with the spin signal reaching zero at the bottom of the SV feature\cite{ciorga2009}. From the measurement we extract the obtained spin injection efficiency $P$ as approximately $51\%$. The discrepancy between this value and the one extracted earlier from Hanle measurements could stem from the anisotropy in $P$, resulting from different orientation of the injected spins with respect to cristallographic directions. For the SV-extracted value we used the signal $\Delta V$ registered when the magnetizations of the magnetic contacts, determining the orientation of injected spins, are parallel to $B_{y}$, i.e., point along [010] direction. Hanle measurements, on the other hand, probe the spins injected parallel to the magnetizations at $B_{y}=0 T$, which do not have to point along [010] at zero field\cite{ciorga2007}.

To investigate the in-plane anisotropy in $P$ we performed measurements while rotating the sample in a magnetic field of 1T, what kept magnetizations of both injector and detector, and also injected spins, parallel to the external field. The measured voltage is then described by Eq.(1) and the observed changes in the signal are due to the dependence of polarization $P$ on the crystallographic direction.\cite{comment1} The typical results of measurements are summarized in Fig. 2. The plotted tunneling anisotropic spin polarization (TASP) for spins forming the angle $\phi$ with [010] direction is defined, in reference to [110], as
\begin{equation}
	TASP_{[110]}(\phi)=100\%\times[P(\phi)-P(45\,^{\circ})]/P(45\,^{\circ})
\end{equation}

\begin{figure}[tbp]
\label{f2}
\begin{center}
\includegraphics[width=0.7\columnwidth,clip]{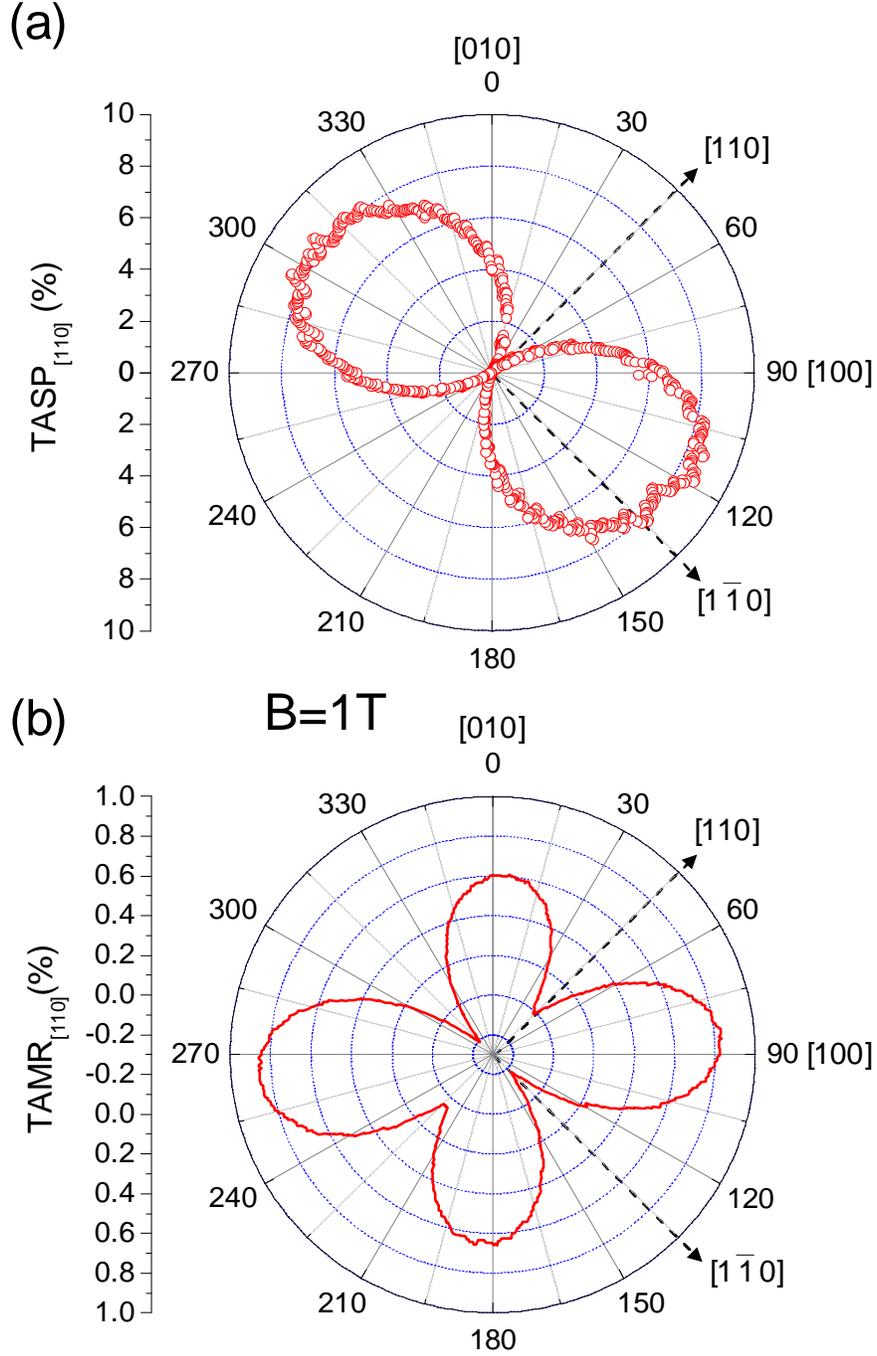}
\caption{(color online)TASP (a) and TAMR (b) data obtained at $B_{in-plane}=1T$; the data was taken while injecting $I_{inj}$ at contacts 3--6 , probing the voltage drop $V_{inj}$ between 3--1 (b) and  detecting the non-local spin signal between 2--1 (a). For details see text.}
\end{center}
\end{figure}
The $\sim8\%$ anisotropy in $P$ is observed between maximum for [1$\bar{1}$0] direction ($\phi=135\,^{\circ}$) and minimum for [110] direction ($\phi=45\,^{\circ}$), which is consistent with theoretical results of Ref.~\onlinecite{sankowski}, assuming a small strain along [110] direction.  In Fig. 2(b) we plot also the anisotropy in tunneling resistance, TAMR, defined in a similar way as TASP, by simply replacing in Eq.(2) spin polarization $P$ with resistance $R=V_{inj}/I_{inj}$. The obtained curve shows anisotropy both along [100] directions as well as along [110] directions, as observed before.\cite{ciorga2007} One can see that the positive $TASP$ along [1$\bar{1}$0], as defined by Eq.2, corresponds to negative TAMR in this direction, the result also being consistent with theoretical results of Ref. \onlinecite{sankowski}. We would like to point out here that we also performed the same type of measurements on other samples using dc measurements and higher bias values and the qualitative behaviour of the signal was always the same, i.e. the spin signal in [1$\bar{1}$0] was always bigger then in direction perpendicular to it, even when the change of sign of TAMR signal was observed upon the switching of the sign of the applied bias\cite{tamr}. 

\begin{figure}
\label{f3}
\begin{center}
\includegraphics[width=0.7\linewidth,clip]{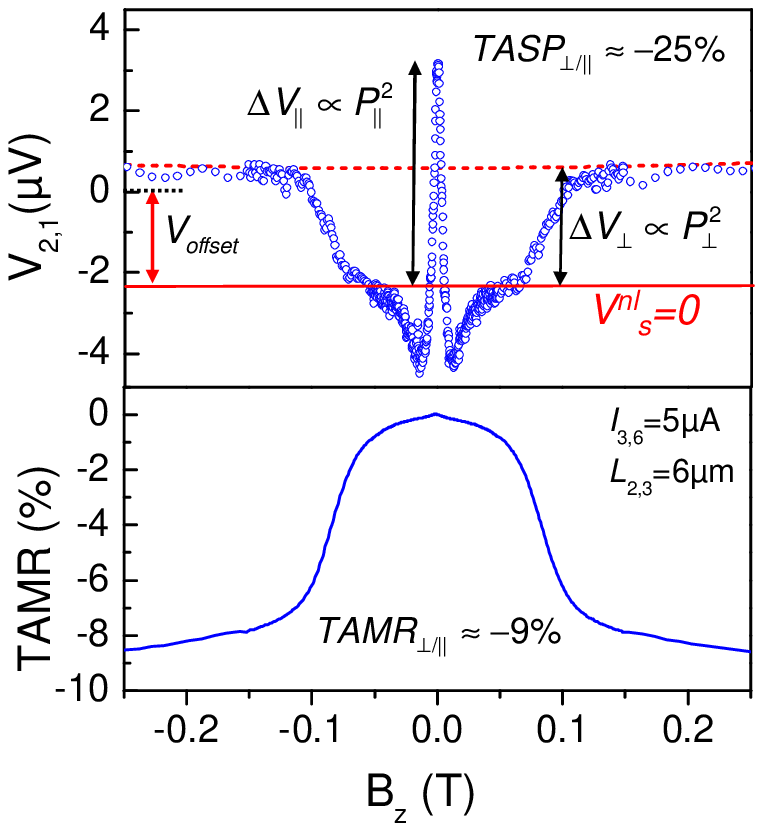}
\caption{(color online) Non-local signal at the detector 2--1 (upper panel) and TAMR of the injector contact 3--6 (lower panel) versus out-of-plane magnetic field. For details see text.}
\end{center}
\end{figure}

Much bigger anisotropy of spin injection efficiency is observed between in-plane and out-of-plane orientation of injected spins. The value of TASP is obtained from measurements in perpendicular magnetic field, presented in Fig. 3. With the spin signal $V^{nl}_{s}$ diminishing with increasing $B$ due to spin dephasing  one can extract an offset signal $V_{offset}\approx -2.2\mu V$, which is bigger than the one in Fig. 1 (different contact pairs are used). A field of value $\sim0.07T$ rotates magnetization of injector and detector out of plane and as a result the injected spins are again parallel to the external field and non-zero spin signal is again observed. The amplitude $\Delta V_{\bot}$ of the resulting step is  proportional to the square of the injection efficiency $P_{\bot}$ of spins pointing along [001]. At higher fields the measured detector signal shows parabolic field dependence that could be attributed to the field dependence of the background.\cite{lou,ciorga2009} This dependence is plotted as a dashed line in the low panel of Fig. 3 and clearly can be disregarded in the plotted range. The difference in the signal registered at high field ($\Delta V_{\bot}$) and the one at 0T ($\Delta V_{\|}$) is then a measure of the anisotropy of the spin injection efficiency. We define perpendicular-to-plane anisotropy as $TASP_{\bot /||}=100\%\times (P_{\bot}-P_{\|})/P_{\|}$ and obtain a value of $\sim-25\%$. In the lower panel of the Fig. 3 we plot a magnetic field dependence of the TAMR signal in reference to the in-plane case, i.e., the value at $B=0T$. The perpendicular-to-plane anisotropy $TAMR_{\bot/||}$ is $\sim-9\%$, the value in the same range as the one in earlier reports on spin Esaki diode contacts\cite{giraud}. The sign of $TASP_{\bot/\|}$ and $TAMR_{\bot/\|}$ is then the same, which is opposite to the in-plane case of $TASP (TAMR)_{[110]}$, discussed earlier.

In summary, we have characterized GaAs-based spin injection devices with $p^{+}-$(Ga,Mn)As/$n^{{+}}-$ GaAs Esaki diode spin aligning contacts for the anisotropy of the obtained spin injection efficiency. The observed in-plane anisotropy for spin oriented either along $[1\bar{1}0]$ or $[110]$ crystallographic direction was found to be $\sim8\%$. The anisotropy between in-plane case and out-of-plane [001] directions was measured as $\sim-25\%$. The positive sign of in-plane anisotropy along $[1\bar{1}0]$ direction suggests the presence of a slight distortion along $[110]$ direction of (Ga,Mn)As contacts.

This work has been supported by Deutsche Forschungs Gemeinschaft (DFG) through Sonderforschungbereich 689.


\begin{thebibliography}{00}
\bibitem{comment1}
We assumed isotropic $V_{offset}$ and the spin diffusion length; the latter was confirmed by analyzing a dependence of the SV-like signal on injector-detector separation length for B-sweeps along different crystallographic directions.
\end{thebibliography}
\end{document}